\documentclass{article}
\usepackage{waspaa21,amsmath,url,times}
\usepackage{color}

\usepackage[dvipdfmx]{graphicx}

\usepackage{amsmath}
\usepackage{amsfonts}
\usepackage{mathrsfs}

\usepackage[ruled,vlined]{algorithm2e}
\SetKwInput{KwInput}{Input}                
\SetKwInput{KwOutput}{Output}              
\SetKwInput{KwInit}{Initialization}

\usepackage{bm,subfig,cite} 

\newcommand{\argmax}{\mathop{\rm arg~max}\limits}
\newcommand{\argmin}{\mathop{\rm arg~min}\limits}
\newcommand{\maximize}{\mathop{\rm maximize}\limits}
\newcommand{\minimize}{\mathop{\rm minimize}\limits}

\title{Kernel learning for sound field estimation\\with $L_1$ and $L_2$ regularizations}

\name{Ryosuke Horiuchi,
      Shoichi Koyama,
      Juliano G. C. Ribeiro,
      Natsuki Ueno,
      Hiroshi Saruwatari                                                                                                                                                   
      }
\address{The University of Tokyo, 7-3-1 Hongo, Bunkyo-ku, Tokyo 113-8656, Japan\\ ryosuke\_horiuchi@ipc.i.u-tokyo.ac.jp         
}

\begin{document}

\ninept
\maketitle

\begin{sloppy}

\begin{abstract}
  
  A method to estimate an acoustic field from discrete microphone measurements is proposed. A kernel-interpolation-based method using the kernel function formulated for sound field interpolation has been used in various applications. The kernel function with directional weighting makes it possible to incorporate prior information on source directions to improve estimation accuracy. However, in prior studies, parameters for directional weighting have been empirically determined. We propose a method to optimize these parameters using observation values, which is particularly useful when prior information on source directions is uncertain. The proposed algorithm is based on discretization of the parameters and representation of the kernel function as a weighted sum of sub-kernels. Two types of regularization for the weights, $L_1$ and $L_2$, are investigated. Experimental results indicate that the proposed method achieves higher estimation accuracy than the method without kernel learning. 
  
\end{abstract}


\vspace{-3pt}

\section{Introduction}
\label{sec:intro}
\vspace{-6pt}

Estimating and interpolating an acoustic field from discrete measurements of microphones are fundamental problems in acoustic signal processing. Such estimations can be applied to the visualization of acoustic fields~\cite{maynard1985nearfield}, interpolation of room impulse responses~\cite{mignot2013room,Ribeiro:IEEE_SAM2020}, identification of sound sources~\cite{park2005sound, teutsch2006acoustic}, capturing sound fields for spatial audio~\cite{poletti2005three,Koyama:IEEE_J_ASLP2013,Iijima:IEEE_MMSP2020}, and spatial active noise control~\cite{zhang2018active,maeno2020spherical}, among others. We focus on the sound field estimation problem in a source-free region.

A typical strategy of sound field estimation is to decompose the measurements into spatial Fourier basis functions~\cite{eWilliams1999}, such as plane waves~\cite{Koyama:IEEE_J_ASLP2013} and spherical harmonics~\cite{laborie2003new, samarasinghe2014wave}. However, the empirical setting of the truncation order and expansion center for the basis expansion is necessary. Sparsity-based approaches using the same basis functions have also been widely investigated~\cite{wabnitz2011upscaling,Koyama:IEEE_J_JSTSP2019} to increase the spatial resolution. The main drawback of this method is that the inference operator of expansion coefficients becomes nonlinear. Thus, the estimation is basically performed by iterative processing. 

The infinite-dimensional analysis of a sound field is proposed in \cite{Ueno:IEEE_SPL2018}, which corresponds to the kernel ridge regression when estimating a pressure field with pressure microphones~\cite{Ueno:IWAENC2018}. This method does not require the empirical setting of truncation order and expansion center. Furthermore, the estimation is performed by a linear operation. In \cite{Ito:ICASSP2020,Ueno:IEEE_J_SP_2021}, the kernel function using prior information on source directions is proposed. The estimation accuracy can be higher than that of the method without prior source direction~\cite{Ito:ICASSP2019,Ueno:IWAENC2018} by using the directionally weighted kernel. 

The kernel function with directional weighting includes two parameters to be set, which are derived from the parameters of the von Mises--Fisher distribution~\cite{mardia2009directional}. One is the prior source directions and the other represents the spread of the weighting. In \cite{Ito:ICASSP2020,Ueno:IEEE_J_SP_2021}, these parameters were empirically determined; however, source directions are not necessarily available in practical situations. Moreover, the optimal setting of the spread parameter is not a trivial task. 

We propose a method to optimize the parameters of the directional kernel function from microphone measurements. We simplify the problem by discretizing the parameters and representing the kernel function as a weighted sum of sub-kernels. As a result, an optimization problem similar to the \textit{multiple kernel learning}~\cite{gonen2011multiple, sonnenburg2006large} is derived. We investigate two types of regularizations for the weighting parameter to derive an algorithm for solving this problem. Although the kernel function is optimized by iterative processing, the estimation process is still a linear operation. We performed numerical simulations in a three-dimensional (3D) space to evaluate our proposed method.

\vspace{-3pt}

\section{PROBLEM STATEMENT AND PRIOR WORKS}
\label{sec:problem}
\vspace{-6pt}

Suppose a region of interest $\Omega \subseteq \mathbb{R}^{3}$ is a simply connected open subset of $\mathbb{R}^3$. The pressure field at the position $\bm{r} \in \Omega$ and angular frequency $\omega\in\mathbb{R}$ is denoted as $u(\bm{r},\omega)$ (i.e., $u:\Omega \times \mathbb{R} \rightarrow \mathbb{C}$). When $\Omega$ does not include any sources, $u$ satisfies the homogeneous Helmholtz equation as
\begin{eqnarray}
\left(\Delta+k^{2}\right) u(\bm{r},\omega) = 0,
\label{eq:helm}
\end{eqnarray}
where $\Delta$ denotes the Laplace operator and $k := \omega/c$ is the wavenumber with the sound velocity $c$. We assume that the pressure field is measured by $M$ omnidirectional microphones arbitrarily placed inside $\Omega$. The position and observed signal of the $m$th microphone are denoted as $\bm{r}_m \in \Omega$ and $s_m(\omega) \in \mathbb{C}$, respectively, ($m\in\{1,\ldots,M\}$). Hereafter, $\omega$ is omitted for notational simplicity.

Our objective is to estimate $u(\bm{r})$ for $\bm{r}\in\Omega$ from the microphone measurements $\{s_m\}_{m=1}^M$ (see Fig.~\ref{fig:problem_setting}). This sound field estimation problem is formulated as follows:
\begin{align}
    \minimize_{u \in \mathscr{H}} \sum_{m=1}^{M} \left| u(\bm{r}_{m})-s_{m}\right|^{2} +\lambda\|u\|_{\mathscr{H}}^{2},
\label{eqn:prior}
\end{align}
where $\mathscr{H}$ is some function space for which we seek a solution, and $\|\cdot\|_{\mathscr{H}}$ is a norm on $\mathscr{H}$. 

When $\mathscr{H}$ is a function space called a \textit{reproducing kernel Hilbert space (RKHS)}, \eqref{eqn:prior} corresponds to kernel ridge regression~\cite{murphy2012machine}. We assume that RKHS $\mathscr{H}$ is defined with the inner product $\langle \cdot, \cdot \rangle_{\mathscr{H}}$ and positive-definite kernel $\kappa : \mathscr{H}\times\mathscr{H}\rightarrow \mathbb{C}$. On the basis of  the representer theorem~\cite{dinuzzo2012representer}, its solution is represented as
\begin{align}
u(\bm{r}) = \sum_{m=1}^M \alpha_m \kappa(\bm{r},\bm{r}_m),
\end{align}
where $\alpha_m\in\mathbb{C}$. Thus, \eqref{eqn:prior} is transformed to the optimization problem for $\alpha_m$ and its solution is obtained as
\begin{align}
\bm{\alpha} = (\bm{K} + \lambda \bm{I})^{-1} \bm{s},
\end{align}
where $\bm{\alpha}=[\alpha_1,\ldots,\alpha_M]^{\mathsf{T}}$, $\bm{s}=[s_1,\ldots,s_M]^{\mathsf{T}}$, and $\bm{K}$ is the Gram matrix defined as
\begin{align}
    \bm{K} &= 
    \begin{bmatrix}
    \kappa(\bm{r}_1,\bm{r}_1) & \cdots & \kappa(\bm{r}_1, \bm{r}_M) \\
    \vdots & \ddots & \vdots \\
    \kappa(\bm{r}_M,\bm{r}_1) & \cdots & \kappa(\bm{r}_M, \bm{r}_M) 
    \end{bmatrix}. 
\end{align}
An appropriate kernel function $\kappa$ associated with $\mathscr{H}$ and $\langle \cdot, \cdot \rangle_{\mathscr{H}}$ should be defined to obtain an appropriate estimate of $u(\bm{r})$.

    \begin{figure}
        \centering
        \includegraphics[width=7cm]{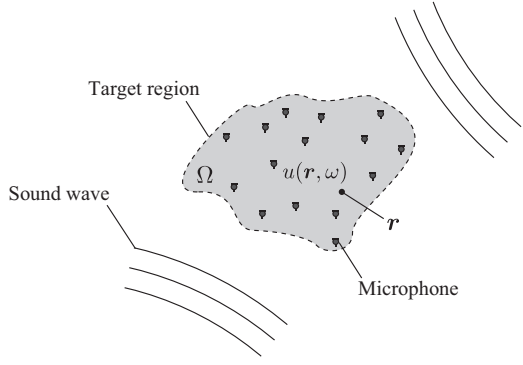}
        \vspace{-4pt}
        \caption{Sound field estimation inside $\Omega$ using discrete set of microphones.}
        \label{fig:problem_setting}
        \vspace{-4pt}
    \end{figure}

\vspace{-3pt}
\subsection{Kernel function for sound field estimation}
\vspace{-3pt}

The pressure field $u$, i.e, the solution of \eqref{eq:helm}, can be well approximated by the superposition of plane waves~\cite{Ueno:IEEE_J_SP_2021}, i.e., the Herglotz wave function~\cite{guevara2018approximation}, as 
\begin{align}
       u(\bm{r}) = \int_{\mathbb{S}_{2}} \tilde{u}(\bm{x}) e^{\mathrm{j} \bm{k}^{\mathsf{T}} \bm{r}} \mathrm{d} \bm{x},
\end{align}
where $\tilde{u}$ is the (square-integrable) complex amplitude of the plane wave of arrival direction ( $\bm{x}\in\mathbb{S}_2$unit sphere in $\mathbb{R}^3$), and $\bm{k}=-k\bm{x}$ is the wave vector. The kernel function using this representation is proposed in \cite{Ito:ICASSP2020,Ueno:IEEE_J_SP_2021}. The inner product and norm over the Hilbert space $\mathscr{H}$ are defined as
\begin{align}
    &\langle u_{1}, u_{2}\rangle_{\mathscr{H}} := \int_{\bm{x} \in \mathbb{S}_{2}} \frac{\tilde{u}_{1}(\bm{x})^{\ast} \tilde{u}_{2}(\bm{x})}{w(\bm{x})} \mathrm{d} \bm{x}, \label{eq:inn_prod}\\
    & \| u \|_{\mathscr{H}} := \sqrt{\langle u, u\rangle_{\mathscr{H}}}, \label{eq:norm}
\end{align}
where $w(\bm{x})$ is a directional weighting function. This function is introduced to incorporate prior knowledge on source directions. The kernel function $\kappa(\bm{r}_1,\bm{r}_2)$ is set as
\begin{align}
    \kappa(\bm{r}_1,\bm{r}_2) = \frac{1}{4\pi} \int_{\mathbb{S}_2} w(\bm{x}) e^{-\mathrm{j}\bm{k}^{\mathsf{T}}(\bm{r}_1-\bm{r}_2)} \mathrm{d}\bm{x}.
    \label{eq:ker_dir}
\end{align}

A specific $w(\bm{x})$ is defined by using the von Mises--Fisher distribution in directional statistics~\cite{mardia2009directional} as
\begin{align}
    & w(\boldsymbol{x}) := \frac{1}{4 \pi C(\beta)} e^{\beta \bm{\eta}^{\mathsf{T}} \bm{x}}, \quad \left(\bm{x} \in \mathbb{S}_{2}\right), \label{eq:weight}\\
    & C(\beta):=
    \begin{cases}
        \displaystyle 1, & \beta=0 \\
        \displaystyle \frac{e^{\beta} - e^{-\beta}}{2\beta}, & \beta \in(0, \infty)
    \end{cases},
\end{align}
where $\beta\in[0,\infty)$ is a constant parameter and $\bm{\eta}\in\mathbb{S}_2$ represents the prior arrival direction of the source. For $\beta>0$, the smaller $w(\bm{x})$ is, the larger the norm $\|u\|_{\mathscr{H}}$ in \eqref{eq:norm} becomes, and vice versa. Thus, the regularization term \eqref{eqn:prior} becomes larger when the difference between the prior arrival direction $\bm{\eta}$ and the direction of $\bm{x}$ becomes larger. On the other hand, when $\beta=0$, the arrival directions of sound waves are assumed to be uniform. By substituting $w(\bm{x})$ in \eqref{eq:weight} into \eqref{eq:ker_dir}, the kernel function with directional weighting is derived as~\cite{Ueno:IEEE_J_SP_2021}
\begin{align}
    & \kappa(\bm{r}_1,\bm{r}_2) = \frac{1}{C(\beta)} j_0 \Big( \left[ (\mathrm{j}\beta\sin\theta\cos\phi-k x_{12})^2 \right. \notag\\
    & \quad \quad \left. + (\mathrm{j}\beta\sin\theta\sin\phi-k y_{12})^2 + (\mathrm{j}\beta\cos\theta-k z_{12})^2  \right]^{\frac{1}{2}} \Big),
    \label{eq:ker_dir_bessel}
\end{align}
where $j_0(\cdot)$ is the $0$th-order spherical Bessel function of the first kind, $\phi$ and $\theta$ are respectively the azimuth and zenith angles of $\bm{\eta}$, and $[x_{12},y_{12},z_{12}]^{\mathsf{T}} := \bm{r}_1-\bm{r}_2$. 

\vspace{-3pt}
\subsection{Problem statement}
\vspace{-3pt}

The weighting function $w(\bm{x})$ defined in \eqref{eq:weight} represents a unimodal distribution directed to a single source. The parameters $\bm{\eta}$ and $\beta$ must be set to specify the kernel function \eqref{eq:ker_dir_bessel}. These parameters are empirically determined using prior information on source directions in previous studies. However, in practice, the prior source directions $\bm{\eta}$ are not necessarily available. Even when $\bm{\eta}$ is accurately given, the appropriate setting of $\beta$ is not obvious. Furthermore, more than one source can exist and reverberation can be non-negligible.  

Our goal is to jointly optimize the weighting function $w$ and the sound field estimate $u$ using the observation values, which can be formulated as
\begin{align}
       \minimize_{u\in\mathscr{H}, w} \sum_{m=1}^{M} \left| u(\bm{r}_m) -  s_{m}\right|^{2} +\lambda \| u \|_{\mathscr{H}}^2.
       \label{eqn:main}
\end{align}
Such a joint optimization is particularly useful when the prior information on the sources is uncertain or the target sound field is complex because of multiple sources and/or reverberation.

\vspace{-3pt}

\section{PROPOSED METHOD}
\vspace{-6pt}

We generalize the weighting function by a convex combination of \eqref{eq:weight} as
\begin{eqnarray}
        w(\bm{x}) = \sum_{d=1}^{D} \frac{\gamma_d}{4 \pi C(\beta_{d})} e^{\beta_d \bm{\eta}_{d}^{\mathsf{T}}\bm{x}},
\end{eqnarray}
where $\gamma_d$ ($d\in\{1,\ldots, D\}$) is the weighting coefficient in $\mathbb{R}_{\ge 0}$, i.e., non-negative real constant. The parameters $\bm{\eta}$ and $\beta$ are discretized into $D_{\bm{\eta}}$ values of $\bm{\eta}_d$ and $D_\beta$ values of $\beta_d$ (i.e., $D=D_{\bm{\eta}}D_\beta$). Then, the kernel function is represented as the weighted sum of the sub-kernels as
\begin{align}
    \kappa(\bm{r}_1, \bm{r}_2) = \sum_{d=1}^D \gamma_d \kappa_d(\bm{r}_1,\bm{r}_2|\bm{\eta}_d,\beta_d) \quad \text{s.t.} \quad \gamma_d \geq 0,
    \label{eq:kernel}
\end{align}
where the sub-kernel $\kappa_d$ is given by \eqref{eq:ker_dir_bessel} with $\bm{\eta}_d$ and $\beta_d$. Thus, problem \eqref{eqn:main} is transformed into the problem of determining $\gamma_d$ as well as $u$. This problem is known as multiple kernel learning~\cite{gonen2011multiple, sonnenburg2006large} in the context of machine learning. We also denote the vector of $\gamma_d$ as $\bm{\gamma}=[\gamma_1,\ldots,\gamma_D]^{\mathsf{T}}\in\mathbb{R}_{\geq 0}^D$.


To jointly optimize $\bm{\gamma}$ and $u$, some constraint should be imposed on $\bm{\gamma}$ because excessively large $\bm{\gamma}$ can lead to a small cost function value. We investigate two types of regularization for $\bm{\gamma}$: $L_1$ and $L_2$.

\vspace{-3pt}
\subsection{Kernel learning with $L_{1}$ regularization}
\vspace{-3pt}

\begin{algorithm}[t]
\DontPrintSemicolon
\SetAlgoLined
 \KwInit{$ \gamma_{d} = 1/D \quad (d = 1, \ldots, D) $}
 \While{\text{stopping criteria are not satisfied}}{
    Compute $J(\bm{\gamma})$\;
    $d_{\max} = \argmax_{d} \gamma_{d}$\;
    Compute $\bm{\delta}$\;
    $\bar{J} = 0, \bar{\bm{\gamma}} = \bm{\gamma}, \bar{\bm{\delta}} = \bm{\delta}$\;
    \While{$\bar{J} < J(\boldsymbol{\gamma})$}{
        $\bm{\gamma} = \bar{\bm{\gamma}}, \bm{\delta} = \bar{\bm{\delta}}$\;
        $\nu = \argmin_{\{d \ | \ \delta_{d} < 0\}} -\gamma_{d}/\delta_{d}, \ \rho_{\max} = -\gamma_{\nu}/\delta_{\nu}$\;
        $\bar{\bm{\gamma}} = \bm{\gamma} + \rho_{\max}\bm{\delta}$\;
        $d_{\max} = \argmax_{d} \gamma_{d}$\;
        $\bar{\delta}_{d_{\max}} = \delta_{d_{\max}} - \delta_{\nu}, \ \delta_{\nu} = 0$\;
        Compute $\bar{J} = J(\bar{\bm{\gamma}})$
    }
    Line search for $\rho \in [0, \rho_{\text{max}}]$ along $\bm{\delta}$\;
    $\bm{\gamma} = \bm{\gamma} + \rho \bm{\delta}$
 }
    \caption{Algorithm for $L_{1}$ regularization}
    \label{algo:l1}
\end{algorithm}

First, we consider the $L_1$ regularization for $\bm{\gamma}$. The optimization problem is written as
\begin{align}
& \minimize_{\bm{\gamma}\in\mathbb{R}_{\geq 0}^D, u\in\mathscr{H}} \sum_{m=1}^M \left| u(\bm{r}_m) - s_m \right|^2 + \lambda \|u\|_{\mathscr{H}}^2 \label{eqn:l1}\\
& \text{s.t.}
\begin{cases}
\displaystyle \kappa(\bm{r},\bm{r}_m) = \sum_{d=1}^D \gamma_d \kappa_d(\bm{r},\bm{r}_m|\bm{\eta}_d,\beta_d) \\
\displaystyle \|\bm{\gamma}\|_1 = \sum_{d=1}^D \gamma_d = 1
\end{cases}. \notag
\end{align}
The constraint on $L_1$-norm can promote sparsity of $\bm{\gamma}$, meaning that a small number of sub-kernels $\kappa_d$ for representing the kernel function $\kappa$ will be selected. Moreover, it can be guaranteed that the directional weighting function $w(\bm{x})$ satisfies $\int_{\mathbb{S}_2} w(\bm{x}) \mathrm{d}\bm{x} = 1$. 

To solve \eqref{eqn:l1}, we apply the algorithm proposed in \cite{rakotomamonjy2008simplemkl}. First, \eqref{eqn:l1} is rewritten as
\begin{eqnarray}
    \minimize_{\bm{\gamma}} J(\bm{\gamma}) \quad \text{s.t.} \quad \|\bm{\gamma}\|_1=1, \ \gamma_d \geq 0,
\label{eqn:l1-main}
\end{eqnarray}
where $J(\bm{\gamma})$ is the solution of the kernel ridge regression as
\begin{eqnarray}
    J(\bm{\gamma}) := \minimize_{\bm{\alpha}} \| \bm{K}\bm{\alpha} - \bm{s} \|^2 + \lambda \bm{\alpha}^{\mathsf{H}} \bm{K} \bm{\alpha}.
    \label{eqn:krr}
\end{eqnarray}
Here, the Gram matrix $\bm{K}\in\mathbb{C}^{M \times M}$ is the weighted sum of the Gram matrices consisting of $\kappa_d$, defined as $\bm{K}^{(d)}\in\mathbb{C}^{M \times M}$, obtained as
\begin{align}
    \bm{K} = \sum_{d=1}^D \gamma_d \bm{K}^{(d)}.
\end{align}
As discussed in Sect.~\ref{sec:problem}, \eqref{eqn:krr} has a closed-form solution. Thus, the gradient of $J(\bm{\gamma})$ with respect to $\gamma_d$ is obtained as
\begin{align}
    \frac{\partial J}{\partial \gamma_{d}} = -\lambda \bm{\alpha}^{\mathsf{H}}\bm{K}^{(d)}\bm{\alpha},
\end{align}
Then, the reduced gradient method is applied to solve \eqref{eqn:l1-main} with the constraints of $\bm{\gamma}$ satisfied. The descent direction $\bm{\delta}\in\mathbb{R}^D$ consisting of $\delta_d$ is derived as
\begin{align}
    \delta_{d} = 
    \begin{cases}
    \displaystyle 0, \quad \quad \text{for} \ \ \gamma_{d} = 0 \ \text{and} \ \frac{\partial J}{\partial \gamma_{d}} -  \frac{\partial J}{\partial \gamma_{d_{\text{max}}}} > 0 \\
    \displaystyle - \frac{\partial J}{\partial \gamma_{d}} +  \frac{\partial J}{\partial \gamma_{d_{\text{max}}}}, \quad \quad \text{for} \ \ \gamma_{d} > 0 \ \text{and} \ d \neq d_{\text{max}} \\ 
    \displaystyle \sum_{d' \neq d_{\text{max}}, \gamma_{d'} > 0} \frac{\partial J}{\partial \gamma_{d'}} -  \frac{\partial J}{\partial \gamma_{d_{\text{max}}}}, \quad \quad \text{for} \ \ d = d_{\text{max}}
    \end{cases}.
\end{align}
Here, $d_{\text{max}}$ is the index of the largest element of $\bm{\gamma}$. The line search of the step-size parameter is also necessary to ensure $\gamma_d \geq 0$. The iteration is repeated until some stopping criteria are satisfied. The $L_1$ regularization algorithm is summarized in Algorithm~\ref{algo:l1}.

\vspace{-3pt}
\subsection{Kernel learning with $L_{2}$ regularization}
\vspace{-3pt}

\begin{algorithm}[t]
\DontPrintSemicolon
\SetAlgoLined
 \KwInit{$\sigma = 0.5, \gamma_{d} = 1/D \quad (d= 1, \ldots, D)$, $\bm{K}=\sum_d \gamma_d \bm{K}^{(d)}$, $\bm{\alpha}=(\bm{K}+\lambda \bm{I})^{-1}$
 }
 \While{stopping criteria are not satisfied}{
  $\bm{v} = \left[\bm{\alpha}^{\top} \bm{K}^{(1)} \bm{\alpha}, \ldots, \bm{\alpha}^{\mathsf{T}} \bm{K}^{(D)} \bm{\alpha}\right]^{\mathsf{T}}$\;
   $\bm{\gamma} = \frac{\bm{v}}{\|\bm{v}\|}$\;
   $\bm{K} = \sum_d \gamma_d \bm{K}^{(d)}$\;
   $\bm{\alpha} = \sigma \bm{\alpha}+(1-\sigma)(\bm{K}+\lambda \bm{I})^{-1}\bm{s}$\;
 }
 \caption{Algorithm for $L_{2}$ regularization}
 \label{alg:l2}
\end{algorithm}

Next, $L_2$ regularization for $\bm{\gamma}$ is considered. The optimization problem is written as
\begin{align}
& \minimize_{\bm{\gamma}\in\mathbb{R}_{\geq 0}^D, u\in\mathscr{H}} \sum_{m=1}^M \left| u(\bm{r}_m) - s_m \right|^2 + \lambda \|u\|_{\mathscr{H}}^2 \label{eqn:l2}\\
& \text{s.t.}
\begin{cases}
\displaystyle \kappa(\bm{r},\bm{r}_m) = \sum_{d=1}^D \gamma_d \kappa_d(\bm{r},\bm{r}_m|\bm{\eta}_d,\beta_d) \\
\displaystyle \|\bm{\gamma}\|_2^2 = \sum_{d=1}^D \gamma_d^2 \leq 1 
\end{cases}. \notag
\end{align}
By the relaxation of the $L_1$-norm equality constraint with the bounded $L_2$-norm, this constraint puts importance on the fitting of the kernel function to the observation values rather than the sparsity of $\bm{\gamma}$. Moreover, the resulting algorithm can be faster than that of the $L_1$ regularization. 

To solve \eqref{eqn:l2}, we apply the algorithm proposed in \cite{cortes2012l2}. The dual problem of the kernel ridge regression \eqref{eqn:prior} is written as
\begin{align}
    \maximize_{\bm{\alpha}\in\mathbb{C}^M} -  \bm{\alpha}^{\mathsf{H}} \bm{\alpha} - \frac{1}{\lambda} \bm{\alpha} \bm{K} \bm{\alpha} +\bm{s}^{\mathsf{H}} \bm{\alpha}  + \bm{\alpha}^{\mathsf{H}} \bm{s}.
\end{align}
Then, problem \eqref{eqn:l2} can be rewritten as
\begin{align}
& \min_{\bm{\gamma}\in\mathbb{R}_{\geq 0}^D} \max_{\bm{\alpha}\in\mathbb{C}^M} -  \bm{\alpha}^{\mathsf{H}} \bm{\alpha} - \frac{1}{\lambda} \sum_{d=1}^{D} \gamma_d \bm{\alpha} \bm{K}^{(d)} \bm{\alpha} + \bm{s}^{\mathsf{H}} \bm{\alpha}  + \bm{\alpha}^{\mathsf{H}} \bm{s} \label{eq:l2_minmax}\\
& \text{s.t.} \quad
\|\bm{\gamma}\|_2^2 = \sum_{d=1}^D \gamma_d^2 \leq 1 .
\notag
\end{align}
By transforming \eqref{eq:l2_minmax} into the max--min problem, the minimization problem for $\bm{\gamma}$ and the maximization problem for $\bm{\alpha}$ are alternately solved in \cite{cortes2012l2}. The algorithm is summarized in Algorithm~\ref{alg:l2}.

\vspace{-3pt}

\section{EXPERIMENTS}
\vspace{-6pt}

\begin{figure}
    \centering
    \includegraphics[width=6.0cm,clip]{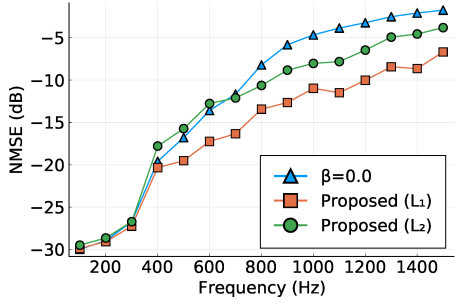}
    \vspace{-4pt}
    \caption{NMSE plotted against frequency}
    \vspace{-4pt}
    \label{fig:nmse_2monopole}
\end{figure}

\begin{figure}
    \centering
    \subfloat[$L_1$ regularization]{\includegraphics[width=120pt,clip]{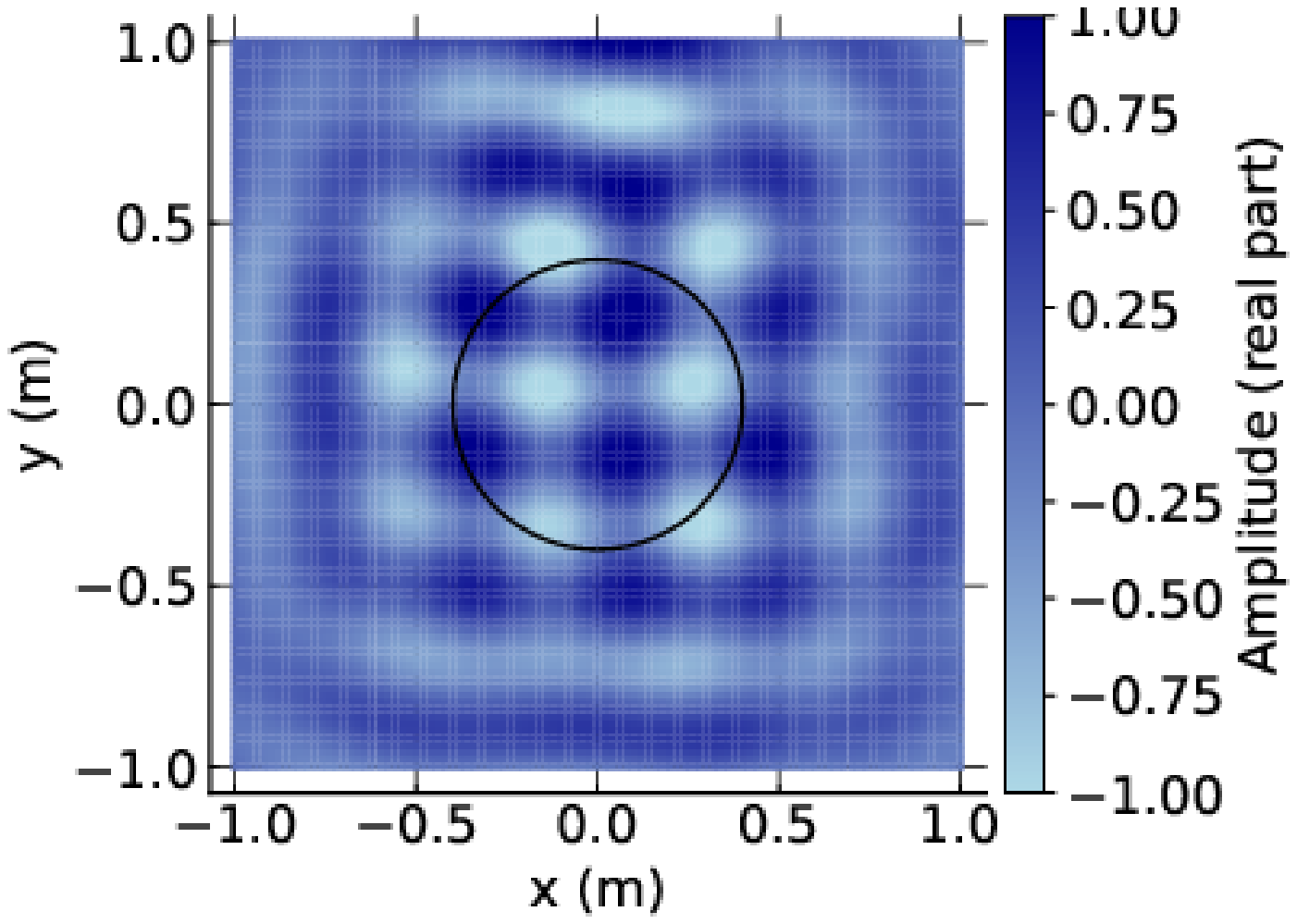}}
    \subfloat[$L_2$ regularization]{\includegraphics[width=120pt,clip]{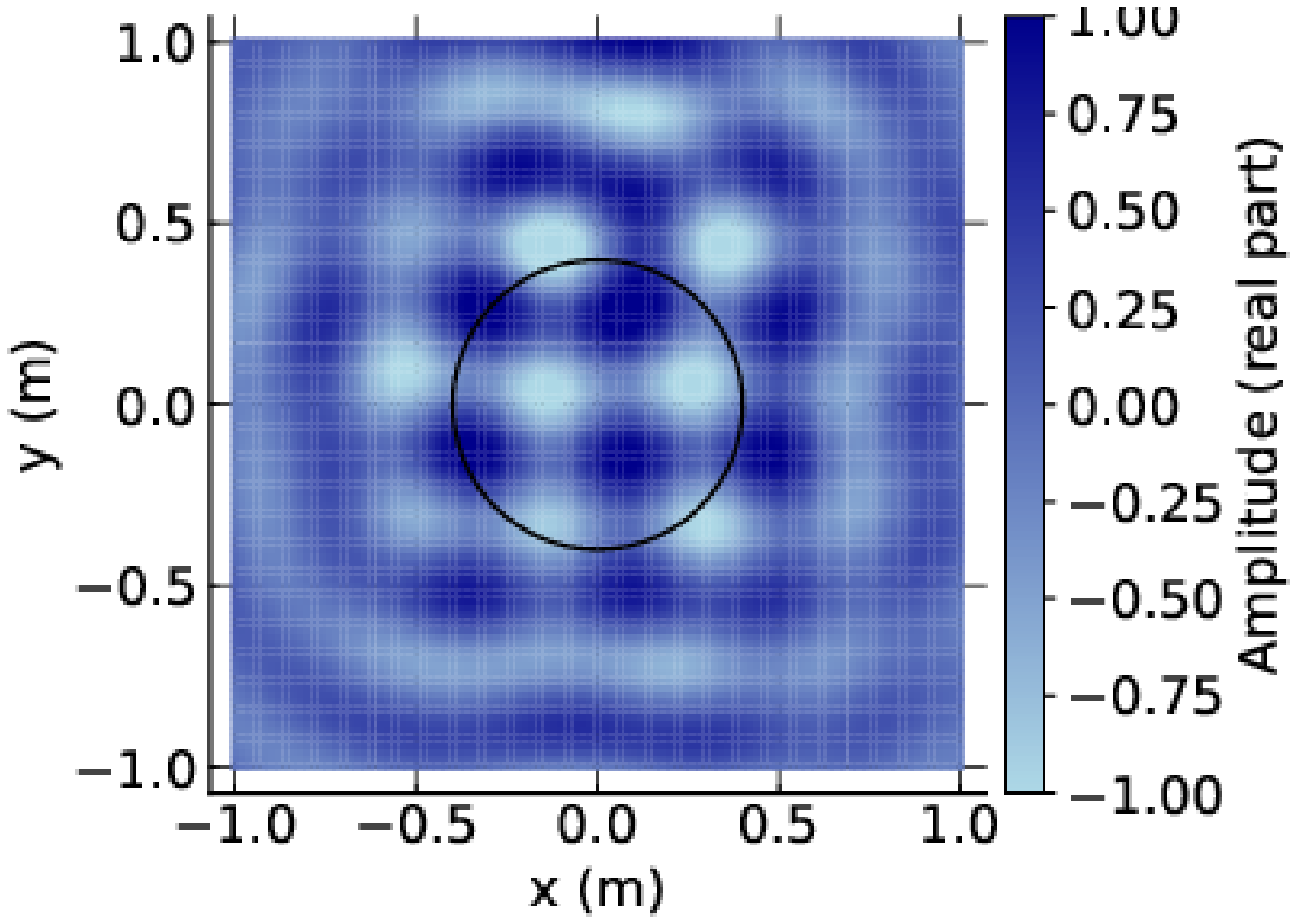}}
    \vspace{-4pt}
    \caption{Estimated pressure distribution}
    \vspace{-4pt}
    \label{fig:exp_pres}
    \centering
    \subfloat[$L_1$ regularization]{\includegraphics[width=120pt,clip]{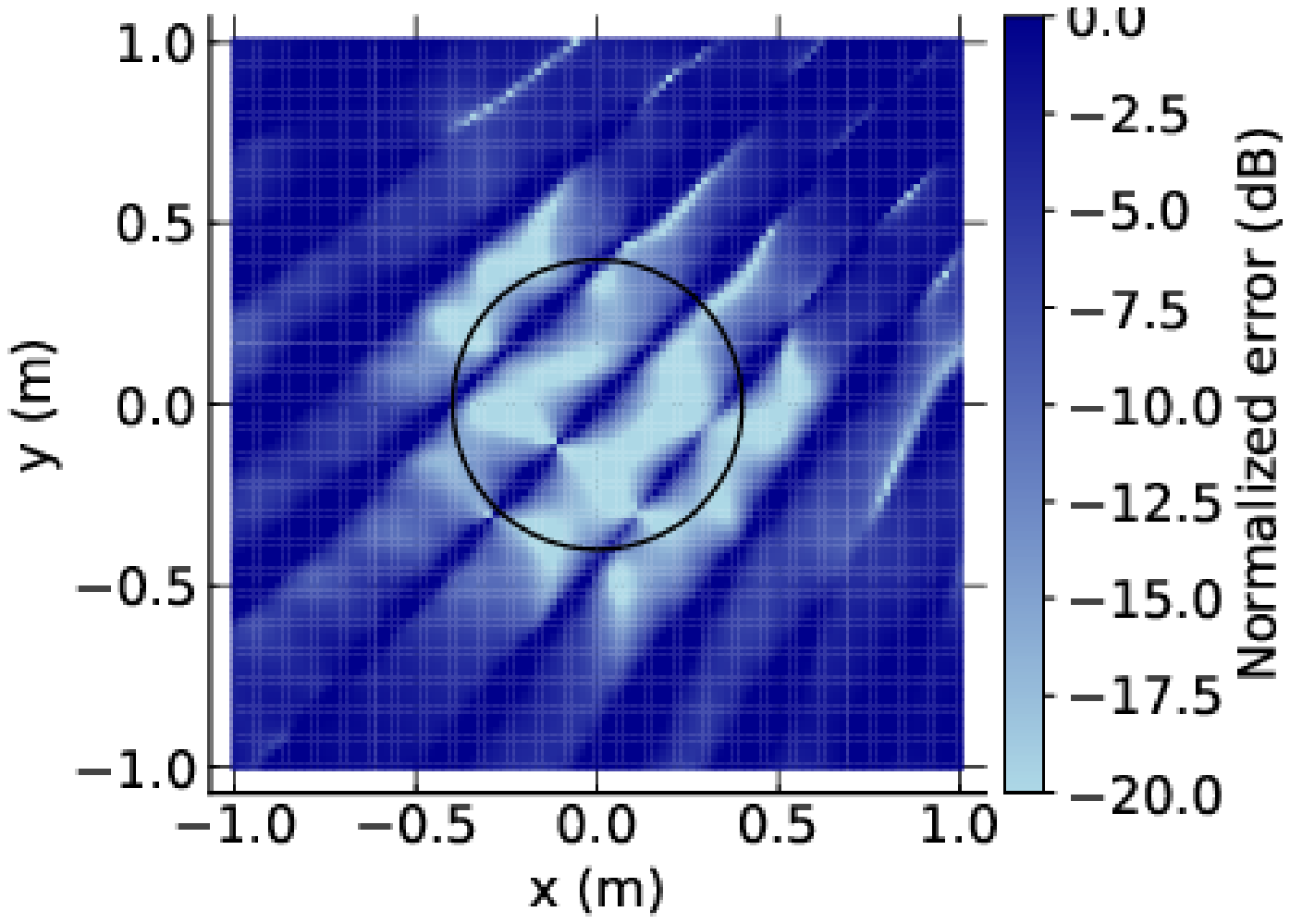}}
    \subfloat[$L_2$ regularization]{\includegraphics[width=120pt,clip]{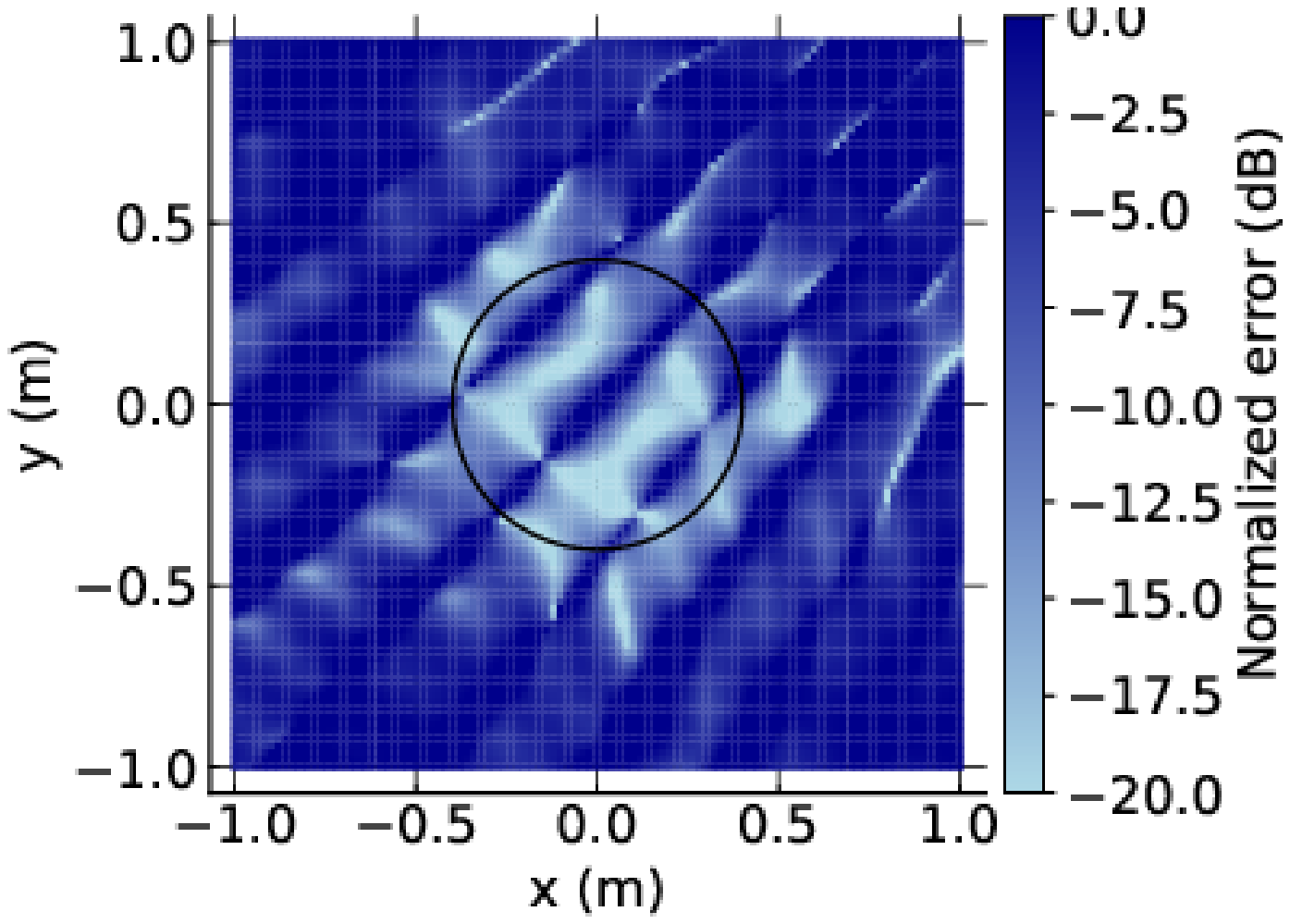}}
    \vspace{-4pt}
    \caption{Normalized error distribution. NMSEs for $L_{1}$ and $L_{2}$ regularization were $-12.64~\mathrm{dB}$ and $-8.82~\mathrm{dB}$, respectively.}
    \vspace{-4pt}
    \label{fig:exp_err}
\end{figure}

We conducted numerical experiments in a 3D free field to evaluate the proposed method. For comparison, the method with the fixed kernel function without prior information of source directions~\cite{Ueno:IWAENC2018} is used. This kernel function corresponds to \eqref{eq:kernel} for $\beta=0$, i.e., uniform directional weighting $w(\bm{x})=1$, which becomes $\kappa(\bm{r}_1,\bm{r}_2)=j_0(k\|\bm{r}_1-\bm{r}_2\|)$. 

The target region $\Omega$ was a sphere of radius $0.40~\mathrm{m}$, and the coordinate origin was set at the center of $\Omega$. Omnidirectional microphones were arranged on two layers of spherical surfaces, whose radii were $0.40~\mathrm{m}$ and $0.45~\mathrm{m}$, to avoid the non-uniqueness problem~\cite{eWilliams1999,poletti2005three}. The number of microphones was 25 for each layer, so the total number $M$ was 50. The microphones were arranged on each layer by using the spherical $t$-design~\cite{tdesign}. Two point sources were placed at $(2.5, 0.0, 0.0)~\mathrm{m}$ and $(0.0, 2.5, 1.0)~\mathrm{m}$, and their amplitudes were both $20$. White Gaussian noise was added to the observation signals so that the signal-to-noise ratio (SNR) becomes $20~\mathrm{dB}$. The speed of sound was set to $340~\mathrm{m/s}$. 

In the proposed method, the kernel functions $\kappa_d$ were obtained by discretizing $\bm{\eta}$ and $\beta$ into $10$ values each, i.e., $D_{\bm{\eta}}=10$ and $D_{\beta}=10$. The arrival directions $\bm{\eta}_d$ were set by regularly dividing the azimuth angle $[-\pi,\pi)$. The parameters $\beta_d$ were set from $0.0$ to $9.0$ at intervals of $1.0$. 

As an evaluation measure, we define the following normalized mean squared error (NMSE):
\begin{eqnarray}
    \mathrm{NMSE} := 10 \log_{10}\frac{\sum_{i\in I} |u_{\mathrm{true}}(\bm{r}_{\mathrm{eval}}^{(i)}) -  u_{\mathrm{est}}(\bm{r}_{\mathrm{eval}}^{(i)} ) |^{2}}{\sum_{i\in I} |u_{\mathrm{true}}(\bm{r}^{(i)}_{\mathrm{eval}} ) |^{2}},
\end{eqnarray}
where $u_{\mathrm{true}}$ and $u_{\mathrm{est}}$ respectively denote the true and estimated pressures, and  $\{\bm{r}_{\mathrm{eval}}^{(i)}\}_{i \in I}$ denotes evaluation positions obtained by discretizing $\Omega$ every $0.05~\mathrm{m}$.

Fig.~\ref{fig:nmse_2monopole} shows the relationship between frequency and NMSE. The NMSEs of the two proposed methods were lower than that of the method using $\beta=0$. In particular, $L_1$ regularization performed better than $L_2$ regularization at all the frequencies. This result indicates that the sparsity constraint on $\bm{\gamma}$ was effective in this setting. On the other hand, the computational time of $L_2$ regularization was almost $10^{-2}$ times that of $L_1$ regularization for estimating $\bm{\gamma}$. As an example, the estimated pressure and normalized error distributions on the $x$-$y$-plane at $900~\mathrm{Hz}$ are shown in Figs.~\ref{fig:exp_pres} and \ref{fig:exp_err}, respectively. In this case, 57\% of $\gamma_d$ became $0$ in the $L_1$ regularization. 

\vspace{-3pt}

\section{CONCLUSION}
\vspace{-6pt}

We proposed a kernel-interpolation-based sound field estimation method with learning kernels. The parameters for directional weighting in the kernel functions have been empirically determined on the basis of  prior information on source directions; however, such information is not necessarily available in practical situations. To optimize these parameters on the basis of the observation values, we formulated the optimization problem by discretizing the parameters and representing the kernel function as the weighted sum of sub-kernels. Two types of regularization, $L_1$ and $L_2$, for the weights were investigated. Although the algorithms for solving these optimization problems are iterative processing, the estimation operation is linear and can be implemented using a finite response filter. The proposed method outperformed the method without the kernel learning in the numerical experiments. 


\vspace{-3pt}

\section{Acknowledgment}
\vspace{-6pt}
This work was supported by JST PRESTO Grant Number JPMJPR18J4.

\vfill

\bibliographystyle{IEEEtran}
\bibliography{str_def_abrv,koyama_en,refs21}

\begin{thebibliography}{10}
\providecommand{\url}[1]{#1}
\def\UrlFont{\rmfamily}
\providecommand{\newblock}{\relax}
\providecommand{\bibinfo}[2]{#2}
\providecommand\BIBentrySTDinterwordspacing{\spaceskip=0pt\relax}
\providecommand\BIBentryALTinterwordstretchfactor{4}
\providecommand\BIBentryALTinterwordspacing{\spaceskip=\fontdimen2\font plus
\BIBentryALTinterwordstretchfactor\fontdimen3\font minus
  \fontdimen4\font\relax}
\providecommand\BIBforeignlanguage[2]{{%
\expandafter\ifx\csname l@#1\endcsname\relax
\typeout{** WARNING: IEEEtran.bst: No hyphenation pattern has been}%
\typeout{** loaded for the language `#1'. Using the pattern for}%
\typeout{** the default language instead.}%
\else
\language=\csname l@#1\endcsname
\fi
#2}}

\bibitem{maynard1985nearfield}
J.~D. Maynard, E.~G. Williams, and Y.~Lee, ``Nearfield acoustic holography: I.
  {Theory} of generalized holography and the development of {NAH},'' \emph{J.
  Acoust. Soc. Amer.}, vol.~78, no.~4, pp. 1395--1413, 1985.

\bibitem{mignot2013room}
R.~{Mignot}, L.~{Daudet}, and F.~{Ollivier}, ``Room reverberation
  reconstruction: Interpolation of the early part using compressed sensing,''
  \emph{{IEEE} Trans. Audio, Speech, Lang. Process.}, vol.~21, no.~11, pp.
  2301--2312, 2013.

\bibitem{Ribeiro:IEEE_SAM2020}
J.~G.~C. Ribeiro, N.~Ueno, S.~Koyama, and H.~Saruwatari, ``Kernel interpolation
  of acoustic transfer function between regions considering reciprocity,'' in
  \emph{Proc. {IEEE} Sensor Array Multichannel Signal Process. Workshop
  ({SAM})}, Jun. 2020.

\bibitem{park2005sound}
M.~Park and B.~Rafaely, ``Sound-field analysis by plane-wave decomposition
  using spherical microphone array,'' \emph{J. Acoust. Soc. Amer.}, vol. 118,
  no.~5, pp. 3094--3103, 2005.

\bibitem{teutsch2006acoustic}
H.~Teutsch and W.~Kellermann, ``Acoustic source detection and localization
  based on wavefield decomposition using circular microphone arrays,'' \emph{J.
  Acoust. Soc. Amer.}, vol. 120, no.~5, pp. 2724--2736, 2006.

\bibitem{poletti2005three}
M.~A. Poletti, ``Three-dimensional surround sound systems based on spherical
  harmonics,'' \emph{J. Audio Eng. Soc.}, vol.~53, no.~11, pp. 1004--1025,
  2005.

\bibitem{Koyama:IEEE_J_ASLP2013}
S.~Koyama, K.~Furuya, Y.~Hiwasaki, and Y.~Haneda, ``Analytical approach to wave
  field reconstruction filtering in spatio-temporal frequency domain,''
  \emph{{IEEE} Trans. Audio, Speech, Lang. Process.}, vol.~21, no.~4, pp.
  685--696, 2013.

\bibitem{Iijima:IEEE_MMSP2020}
N.~Iijima, S.~Koyama, and H.~Saruwatari, ``Binaural rendering from distributed
  microphone signals considering loudspeaker distance in measurements,'' in
  \emph{Proc. {IEEE} Int. Workshop Multimedia Signal Process. ({MMSP})},
  Tampere, Sep. 2020.

\bibitem{zhang2018active}
J.~{Zhang}, T.~D. {Abhayapala}, W.~{Zhang}, P.~N. {Samarasinghe}, and
  S.~{Jiang}, ``Active noise control over space: A wave domain approach,''
  \emph{{IEEE/ACM} Trans. Audio, Speech, Lang. Process.}, vol.~26, no.~4, pp.
  774--786, 2018.

\bibitem{maeno2020spherical}
Y.~{Maeno}, Y.~{Mitsufuji}, P.~N. {Samarasinghe}, N.~{Murata}, and T.~D.
  {Abhayapala}, ``Spherical-harmonic-domain feedforward active noise control
  using sparse decomposition of reference signals from distributed sensor
  arrays,'' \emph{{IEEE/ACM} Trans. Audio, Speech, Lang. Process.}, vol.~28,
  pp. 656--670, 2020.

\bibitem{eWilliams1999}
E.~Williams, \emph{Fourier Acoustics: Sound Radiation and Nearfield Acoustic
  Holography}.\hskip 1em plus 0.5em minus 0.4em\relax London, UK: Academic
  Press, 1999.

\bibitem{laborie2003new}
A.~Laborie, R.~Bruno, and S.~Montoya, ``A new comprehensive approach of
  surround sound recording,'' in \emph{Proc. 114th {AES} Conv.}, Amsterdam,
  2003.

\bibitem{samarasinghe2014wave}
P.~{Samarasinghe}, T.~{Abhayapala}, and M.~{Poletti}, ``Wavefield analysis over
  large areas using distributed higher order microphones,'' \emph{{IEEE/ACM}
  Trans. Audio, Speech, Lang. Process.}, vol.~22, no.~3, pp. 647--658, 2014.

\bibitem{wabnitz2011upscaling}
A.~Wabnitz, N.~Epain, A.~McEwan, and C.~Jin, ``Upscaling ambisonic sound scenes
  using compressed sensing techniques,'' in \emph{Proc. {IEEE} Int. Workshop
  Appl. Signal Process. Audio Acoust. ({WASPAA})}, New Paltz, 2011, pp. 1--4.

\bibitem{Koyama:IEEE_J_JSTSP2019}
S.~Koyama and L.~Daudet, ``Sparse representation of a spatial sound field in a
  reverberant environment,'' \emph{{IEEE} J. Sel. Topics Signal Process.},
  vol.~13, no.~1, pp. 172--184, 2019.

\bibitem{Ueno:IEEE_SPL2018}
N.~Ueno, S.~Koyama, and H.~Saruwatari, ``Sound field recording using
  distributed microphones based on harmonic analysis of infinite order,''
  \emph{{IEEE} Signal Process. Lett.}, vol.~25, no.~1, pp. 135--139, 2018.

\bibitem{Ueno:IWAENC2018}
------, ``Kernel ridge regression with constraint of helmholtz equation for
  sound field interpolation,'' in \emph{Proc. Int. Workshop Acoust. Signal
  Enhancement ({IWAENC})}, Tokyo, Sep. 2018, pp. 436--440.

\bibitem{Ito:ICASSP2020}
H.~Ito, S.~Koyama, N.~Ueno, and H.~Saruwatari, ``Spatial active noise control
  based on kernel interpolation with directional weighting,'' in \emph{Proc.
  {IEEE} Int. Conf. Acoust., Speech, Signal Process. ({ICASSP})}, May 2020, pp.
  8399--8403.

\bibitem{Ueno:IEEE_J_SP_2021}
N.~Ueno, S.~Koyama, and H.~Saruwatari, ``Directionally weighted wave field
  estimation exploiting prior information on source direction,'' \emph{{IEEE}
  Trans. Signal Process.}, 2021, (in press).

\bibitem{Ito:ICASSP2019}
H.~Ito, S.~Koyama, N.~Ueno, and H.~Saruwatari, ``Feedforward spatial active
  noise control based on kernel interpolation of sound field,'' in \emph{Proc.
  {IEEE} Int. Conf. Acoust., Speech, Signal Process. ({ICASSP})}, Brighton, May
  2019, pp. 511--515.

\bibitem{mardia2009directional}
K.~V. Mardia and P.~E. Jupp, \emph{Directional Statistics}.\hskip 1em plus
  0.5em minus 0.4em\relax Chichester: John Wiley \& Sons, 2009, vol. 494.

\bibitem{gonen2011multiple}
M.~G{\"o}nen and E.~Alpayd{\i}n, ``Multiple kernel learning algorithms,''
  \emph{J. Mach. Learn. Res.}, vol.~12, pp. 2211--2268, 2011.

\bibitem{sonnenburg2006large}
S.~Sonnenburg, G.~R{\"a}tsch, C.~Sch{\"a}fer, and B.~Sch{\"o}lkopf, ``Large
  scale multiple kernel learning,'' \emph{J. Mach. Learn. Res.}, vol.~7, pp.
  1531--1565, 2006.

\bibitem{murphy2012machine}
K.~P. Murphy, \emph{Machine Learning: A Probabilistic Perspective}.\hskip 1em
  plus 0.5em minus 0.4em\relax Cambridge: MIT Press, 2012.

\bibitem{dinuzzo2012representer}
F.~Dinuzzo and B.~Sch\"{o}lkopf, ``The representer theorem for {Hilbert}
  spaces: a necessary and sufficient condition,'' in \emph{Proc. Adv. Neural
  Inform. Process. Systems (NeurIPS)}, vol.~25, 2012.

\bibitem{guevara2018approximation}
F.~Guevara~Vasquez and C.~Mauck, ``Approximation by {Herglotz} wave
  functions,'' \emph{{SIAM} J. Appl. Math.}, vol.~78, no.~3, pp. 1283--1299,
  2018.

\bibitem{rakotomamonjy2008simplemkl}
A.~Rakotomamonjy, F.~Bach, S.~Canu, and Y.~Grandvalet, ``Simplemkl,'' \emph{J.
  Mach. Learn. Res.}, vol.~9, pp. 2491--2521, 2008.

\bibitem{cortes2012l2}
C.~Cortes, M.~Mohri, and A.~Rostamizadeh, ``L2 regularization for learning
  kernels,'' in \emph{Proc. Conf. Uncertainty in Artificial Intelligence
  (UAI)}, Arlington, 2009, p. 109^^e2^^80^^93116.

\bibitem{tdesign}
X.~Chen and R.~S. Womersley, ``Existence of solutions to systems of
  underdetermined equations and spherical designs,'' \emph{SIAM J. Numer.
  Anal.}, vol.~44, no.~6, pp. 2326--2341, 2006.

\end{thebibliography}
%
%
%
%
%
%
%
%
%

\end{sloppy}
\end{document}